\documentclass[aps,prd,preprint,superscriptaddress,nofootinbib]{revtex4-1}

\usepackage{amsmath}
\usepackage{braket}
\usepackage{bbm}
\usepackage{color}
\usepackage{hyperref}
\usepackage{graphicx}
\DeclareMathOperator{\Tr}{Tr}
\DeclareMathOperator{\Det}{Det}

\begin{document}

\title{Backreaction of particle production on false vacuum decay}

\author{Cyril Lagger}
\email[]{cyril.lagger@sydney.edu.au}

\affiliation{ARC Centre of Excellence for Particle Physics at the Terascale, School of Physics, The University of Sydney, NSW 2006, Australia}

\affiliation{Institut de Th\'{e}orie des Ph\'{e}nom\`{e}nes Physiques, \'{E}cole Polytechnique F\'{e}d\'{e}rale de Lausanne, CH-1015, Lausanne, Switzerland}

\date{\today}

\begin{abstract}
As originally described by Rubakov, particles are produced during the tunneling of a metastable quantum field. We propose to extend his formalism to compute the backreaction of these particles on the semiclassical decay probability of the field. The idea is to integrate out the external bath of particles by computing the reduced density matrix of the system. Following this approach, we derive an explicit correction factor in the specific case of scalar particle production in flat spacetime. In this given framework, we conclude that the backreaction is ultraviolet finite and enhances the decay rate. Moreover, in the weak production limit, the backreaction factor is directly given by one half of the total number of created particles. In order to estimate the importance of this correction, we apply our formalism to a toy model potential which allows us to consider both the decay of a homogeneous bounce and the nucleation of a thin-wall bubble. In the former case, the impact of the created particles is parameter dependent and we exhibit a reasonable choice of variables for which ones the backreaction is significant. In the latter case, we conclude that the backreaction is always negligible.
\end{abstract}


\maketitle

\section{Introduction}
A quantum field lying at a local minimum of its potential has a nonzero probability to tunnel to a more energetically favorable state through barrier penetration. Such a phase transition known as false vacuum decay is typically described as the nucleation of a bubble of true vacuum surrounded by the initial metastable phase \cite{Kobzarev:1974,Coleman:1977}.

This process may occur in various quantum field theories. Currently, a lot of attention is given to the stability of the electroweak vacuum of the Standard Model (SM). Since the discovery of the Higgs boson \cite{ATLAS:2012,CMS:2012} with a mass of $m_H=125.09\pm 0.24$ GeV \cite{Aad:2015}, all the parameters of this theory are known without any sign of new physics. Therefore, extrapolating the SM to high energies, state-of-the-art calculations  indicate that the Universe is lying at the edge between stability and metastability \cite{Degrassi:2012,Bezrukov:2012,Spencer-Smith:2014,Buttazzo:2013,Bednyakov:2015}. In the latter case, the Higgs potential becomes negative at an instability scale around $\Lambda_I \approx 10^{11}$ GeV and thus develops a global minimum at large field values. In this context, the decay probability of the electroweak vacuum may become a crucial parameter to probe both the fate and the history of our Universe. In particular, it has to agree with the observation that the Higgs field has not yet decayed during the early stage of the Universe or during its subsequent evolution.

The usual starting point to evaluate such a false vacuum decay rate in flat spacetime is the instanton method of Coleman \cite{Coleman:1977}. In this formalism, the decay probability per unit time per unit volume is computed at the lowest order of the semiclassical approximation: $\Gamma/V \approx e^{-S_B}$, where $S_B$ is the Euclidean action evaluated along the bounce trajectory. The precision of this result can then be improved including one-loop quantum corrections \cite{Callan:1977} or finite temperature \cite{Affleck:1980,Linde:1981}. When gravitational effects cannot be neglected, e.g. during the early Universe, it is well known that the decay can take place either through the Coleman-de Luccia instanton \cite{Coleman:1980}, which is the analogue of bubble nucleation in flat spacetime, or through the Hawking-Moss instanton \cite{Hawking:1981} which is a homogeneous process allowed when the geometry of the Universe is closed.

These formalisms have been applied to the SM stability in the past few years. They indicate that the lifetime of the current electroweak vacuum is much longer than the age of the Universe by many orders of magnitude, both at the semiclassical \cite{Buttazzo:2013} and one-loop \cite{Isidori:2001,Branchina:2014} level. Thus, it does not contradict any observation. However, the situation is more intriguing during the early Universe assuming a period of inflation. Recent investigations suggest that the survival of the electroweak vacuum during this epoch implies stringent constraints on the Hubble rate and the top quark mass depending on the details of the process, like for example the coupling of the Higgs field to gravity \cite{Kobakhidze:2013,Kobakhidze:2014x,Hook:2014,Kearney:2015,Fairbairn:2014,Enqvist:2014,Herranen:2014,Herranen:2015,Shkerin:2015,Espinosa:2015}. It also appears that the temperature of the Universe during the reheating process following inflation influences the stability of the SM \cite{Espinosa:2015,Espinosa:2007qp,Rose:2015}. In addition to this, some other effects that could change our knowledge of the Higgs decay rate have also been considered including modifications from Planck scale higher-dimensional operators \cite{Branchina:2014,Branchina:2013,Branchina:2014usa,Branchina:2015}, the presence of an impurity in the metastable phase \cite{Grinstein:2015} or the influence of black holes as nucleation seeds \cite{Burda:2015,Burda:2015yfa}.

This paper investigates another effect which has not yet been taken into account to the best of our knowledge and which occurs in every type of false vacuum decay: the backreaction of particle production. In the 1980s, Rubakov \cite{Rubakov:1984, Lavrelashvili:1978} proved that the fluctuations generated during the tunneling of a metastable field produce a spectrum of real particles. This observation was then confirmed by other groups using different formalisms \cite{Vachaspati:1991,Tanaka:1993a}. We propose to review and extend Rubakov's work to evaluate the impact of these particles on the decay rate of the field. Our approach will follow the reduced density matrix formalism used in quantum mechanics to address the influence of the environment on a given quantum process \cite{Landau:1927,Joos:1984,Zurek:2003,Schlosshauer:2003}.\footnote{This formalism was actually proposed to describe the environment-induced decoherence which attempts to explain why the world looks classical despite its quantum nature.} In this scheme, the information related to the system is extracted by integrating out the external degrees of freedom. Applied to our case, the decaying field is considered as the system of interest while the environment corresponds to the external bath of created particles.

It is worth mentioning that the objective of this article is to present our formalism and thus we restrict our attention to the simplest scenario, namely the backreaction of scalar particles in flat spacetime. We postpone the study of spinor and vector fields, necessary to fully address the SM, as well as gravitational effects to a subsequent work. The article is organized as follows. In Sec. \ref{sec:review}, we review the instanton formalism in flat spacetime and the process of particle production as described by Rubakov. In Sec. \ref{sec:backreaction}, we explicitly compute the reduced density matrix for a system of two scalar fields and we derive a correction factor which accounts for the backreaction of particle production. An interpretation of this new factor is given. Section \ref{sec:toy_model} is dedicated to the application of the previous formalism to a toy model potential for which we consider two types of decay: the nucleation of a thin-wall bubble and a homogeneous process. In the last section, we summarize our results and describe what would be our future investigations in particular the application of this formalism to the Higgs potential of the SM.


\section{\label{sec:review}Vacuum decay and particle production}
\subsection{\label{sub:decay_rate}Semiclassical decay rate}
Consider a real scalar field $\sigma$ in Minkowski spacetime:
\begin{equation}
\label{equ:lagrangian_sigma}
\mathcal{L}_{\sigma}=\frac12 \partial_{\mu} \sigma \partial^{\mu} \sigma - V(\sigma).
\end{equation}
We assume that the potential $V$ has a local minimum at $\sigma_F$ (false vacuum) and a global minimum at $\sigma_T$ (true vacuum). In the semiclassical approximation, the decay probability per unit time per unit volume of a field initially trapped near $\sigma_F$ is given by \cite{Coleman:1977}\footnote{Except in Eq. (\ref{equ:semiclassical_decay_rate}), we shall work in natural units $\hbar=c=1$ throughout this paper. }
\begin{equation}
\label{equ:semiclassical_decay_rate}
\frac{\Gamma}{\mathcal{V}}= A \, e^{-\frac{S_B}{\hbar}} \left(1 + O(\hbar) \right).
\end{equation}
The semiclassical exponent $S_B$ corresponds to the Euclidean action $S_E$ computed along the bounce trajectory: $S_E=\int d\tau d^3\mathbf{x}\left[\frac12 \left(\partial_{\tau}\sigma \right)^2 + \frac12 \left(\partial_i\sigma \right)^2 + V(\sigma)\right]$, where we have introduced the formal Euclidean time $\tau=i t$. The bounce $\sigma_B(\mathbf{x}, \tau)$ is the Euclidean trajectory which minimizes $S_E$ with the conditions that it starts from $\sigma_F$ at $\tau=-\infty$, evolves under the barrier until it reaches the boundary with the classically allowed region at $\tau=0$ and then bounces back to $\sigma_F$ at $\tau=+\infty$. In equations, it satisfies
\begin{equation}
\label{equ:bounce_equation_general}
(\partial^2_{\tau}+\partial^2_{i} ) \sigma = \frac{\partial V}{ \partial \sigma}, \quad \sigma(\tau,\mathbf{x}) \xrightarrow{\tau \rightarrow \pm \infty} \sigma_{F}, \quad \left.\frac{\partial \sigma}{\partial \tau}\right|_{(\tau=0,\, \mathbf{x})}=0.
\end{equation}

According to Eq. (\ref{equ:semiclassical_decay_rate}), the decay process is exponentially suppressed for large values of $S_B$. It means that the field cannot decay homogeneously in an open universe. This restricts our attention to two cases of interest: a homogeneous decay taking place in a finite volume $\mathcal{V}$ or the nucleation of a bubble of true vacuum $\sigma_T$. In the former case, Eq. (\ref{equ:bounce_equation_general}) reduces to $\partial^2_{\tau} \sigma = \frac{\partial V}{\partial \sigma}$ with the same boundary conditions. The Euclidean action also simplifies to
\begin{equation}
\label{equ:euclidean_action_homogeneous}
S_{E,hom} = \mathcal{V} \int d\tau \left[\frac12 \left(\partial_{\tau}\sigma \right)^2 + V(\sigma)\right].
\end{equation}
 In the latter case, Coleman proved that it always exists an $O(4)$ symmetric solution $\sigma_B(\rho)$, where $\rho = \sqrt{\tau^2 +\mathbf{x}^2}$, satisfying\footnote{Note that the last condition in Eq. (\ref{equ:bounce_equation_o4}) is not a consequence of (\ref{equ:bounce_equation_general}) but is added to avoid any singular solution at the origin.}
\begin{equation}
\label{equ:bounce_equation_o4}
\frac{d^2\sigma}{d \rho^2}+\frac3{\rho} \frac{d\sigma}{d \rho}=\frac{\partial V}{\partial \sigma}, \quad \sigma(\rho \rightarrow +\infty) = \sigma_{F}, \quad \left.\frac{d \sigma}{d \rho}\right|_{\rho=0}=0.
\end{equation}
The Euclidean action then becomes
\begin{equation}
\label{equ:euclidean_action_o4_bubble}
S_{E,O(4)} = 2 \pi^2 \int_0^{+\infty} d\rho \rho^3 \left[ \frac12 \left( \frac{d \sigma}{d \rho}\right)^2 + V(\sigma) \right].
\end{equation}

The previous equations give all the information to compute the decay rate at the lowest order of the semiclassical approximation. On the other hand, the prefactor $A$ in Eq. (\ref{equ:semiclassical_decay_rate}) corresponds to the one-loop quantum corrections. Using a path integral formulation, Callan and Coleman showed in \cite{Callan:1977} that it reduces to the computation of a functional determinant. In practice, this is often a very difficult task since this quantity is ultraviolet divergent \textit{per se} and requires renormalization techniques. Physically, this prefactor corresponds to the smallest fluctuations of the tunneling field around the bounce trajectory. Thus, roughly speaking, it represents the effect of virtual particles on the decay rate. Our aim in this paper is actually to derive a new correction factor related to the production of real particles during the false vacuum decay.


\subsection{\label{sub:particle_production}Particle production}
Formally, the phenomenon of particle creation takes place once we add another field to the Lagrangian (\ref{equ:lagrangian_sigma}). In this article, we restrict our attention to an external scalar field $\phi$ coupled to $\sigma$ in the following way:
\begin{equation}
\label{equ:lagrangian_total}
\mathcal{L}_{tot}= \mathcal{L}_{\sigma} + \underbrace{\frac12 \partial_{\mu} \phi \partial^{\mu} \phi - \frac12 m^2(\sigma) \phi^2}_{\mathcal{L}_{\phi}},
\end{equation}
where $m^2(\sigma)$ is an arbitrary coupling parameter. It is worth mentioning that this model allows us to consider the self-excitation of the tunneling field around its bounce solution if we write $\sigma=\sigma_B + \phi$ and if $m^2(\sigma)$ is replaced by $V''(\sigma_B)$.

We now review the formalism of Rubakov \cite{Rubakov:1984} who proved that some $\phi$ particles are created during the decay process of $\sigma$. The state vector of the total system $\ket{\Psi}$ satisfies the stationary Schr\"{o}dinger equation
\begin{equation}
\label{equ:schrodinger_equation_hamiltonian_form}
\left( H_{\sigma} + H_{\phi}\right) \ket{\Psi} =E \ket{\Psi},
\end{equation}
where $H_{\sigma}$ and  $H_{\phi}$ are the Hamiltonians associated to the Lagrangians $\mathcal{L}_{\sigma}$ and $\mathcal{L}_{\phi}$. In the absence of $\phi$, the model is well approximated by the quasiclassical bounce trajectory $\sigma_B(\mathbf{x},\tau)$ described in the previous section. In that case, the wave functional associated to the state $\ket{\Psi}$ becomes parametrized, in the field representation, by the Euclidean time $\tau$ as $\braket{\sigma_B(\tau)|\Psi}=\Psi[\sigma_B(\tau)]\approx e^{-S_E[\sigma_B(\tau)]}$ where 
\begin{equation}
S_E[\sigma_B(\tau)]=\int_{-\infty}^{\tau} d\tau' \int d^3\mathbf{x}\left[\frac12 \left(\partial_{\tau'}\sigma_B \right)^2 + \frac12 \left(\partial_i\sigma_B \right)^2 + V(\sigma_B)\right].
\end{equation} 
Then, we assume that the addition of the new field $\phi$ does not significantly change this behavior. In other words, we study the field $\phi$ in the background of the semiclassical tunneling bounce $\sigma_B$ without taking into account the action of the former on the latter. This is actually the purpose of Sec. \ref{sec:backreaction} to investigate this effect. Under this assumption, we can parametrize the total state for the two fields as
\begin{equation}
\label{equ:total_state_separation_bounce}
\ket{\Psi(\tau)} \approx e^{-S_E(\tau)}\ket{\phi(\tau)}, 
\end{equation}
where the state $\ket{\phi(\tau)}:=\ket{\phi[\sigma_B(\tau)]}$ encodes all the information related to the fluctuating field $\phi$ along the Euclidean trajectory [for convenience we simply wrote the dependence in $\tau$ instead of $\sigma_B(\tau)$]. Before tunneling, we ask for this state to be an eigenstate of $H_{\phi}$ according to
\begin{equation}
\label{equ:boundary_condition_phi_state}
H_{\phi}[\sigma_F]\ket{\phi[\sigma_F]}=\mathcal{E} \ket{\phi[\sigma_F]},
\end{equation}
where the Hamiltonian is also parametrized in function of the bounce:
\begin{equation}
\label{equ:hamiltonian_phi_with_bounce}
H_{\phi}[\sigma_B(\tau)]=\int d^3 \mathbf{x} \left[  \frac12 \pi^2_{\phi} +\frac12 \partial_i \phi \partial_i \phi +\frac12 m^2\left( \sigma_B(\mathbf{x}, \tau)\right) \phi^2\right].
\end{equation}
Even in the Schr\"{o}dinger picture (with time-independent operators $\phi(\mathbf{x})$ and $\pi_{\phi}(\mathbf{x})$), we note that the Hamiltonian has an explicit $\tau$ dependence coming from $m^2(\mathbf{x},\tau):=m^2 \left(\sigma_B(\mathbf{x}, \tau)\right)$. This is a typical signature of systems exhibiting particle production like the well-known examples of particle creation by an external electromagnetic field \cite{Schwinger:1951} or by a curved spacetime background \cite{Parker:1969} (with the important difference that our system evolves in Euclidean time rather than in physical time). 

From the above considerations, Rubakov showed that the stationary equation (\ref{equ:schrodinger_equation_hamiltonian_form}) with the boundary condition (\ref{equ:boundary_condition_phi_state}) reduces to a Euclidean Schr\"{o}dinger equation for $\phi$:
\begin{equation}
\label{equ:schrodinger_equation_fluctuating_field}
\frac{\partial \ket{\phi(\tau)}}{\partial \tau}= - (H_{\phi}(\tau)-\mathcal{E}) \ket{\phi(\tau)}.
\end{equation}
Since in this article we are only concerned with $\phi$ initially in its vacuum state, we set $\mathcal{E}=0$ and $\ket{\phi[\sigma_F]}=\ket{O_{\tau_{-}}}$.\footnote{For the interesting discussion of an initially excited state, we refer to Rubakov's original paper \cite{Rubakov:1984}.} For convenience, we write the initial instant as $\tau_-$ remembering that the limit $\tau_{-} \to -\infty$ has to be taken at the end of the calculation. As usual in the Schr\"{o}dinger picture, the solution of the previous equation can be expressed from the operator of evolution 
\begin{equation}
 \label{equ:operator_evolution}
\mathcal{U}_{\tau}= T \exp \left( - \int_{\tau_{-}}^{\tau} H_{\phi}(\tau')d\tau' \right)
\end{equation}
as
\begin{equation}
 \label{equ:solution_with_operator_evolution}
\ket{\phi(\tau)}=\mathcal{U}_{\tau} \ket{O_{\tau_-}},
\end{equation}
where $T$ is the time ordering operator. In contrast with quantum field theory in Minkowski spacetime, this operator is not unitary: $\mathcal{U}_{\tau}^{-1}\neq \mathcal{U}_{\tau}^{\dagger}$ (note the absence of the imaginary factor $i$ in the exponent since we are working in Euclidean time).

We now introduce the set of creation-annihilation operators $b_{\alpha}^{\dagger}$, $b_{\alpha}$ which diagonalize the initial Hamiltonian $H_{\phi}(\tau_{-})$ and satisfy $b_{\alpha}\ket{O_{\tau_-}}=0$ ($\alpha$ represents a generic set of quantum numbers and $\sum_{\alpha}$ shall either refer to summation or integration). Since $H_{\phi}$ is explicitly time dependent, the vacuum state is not conserved. In other words, at a given instant $\tau$, the vacuum is defined by a new set of operators $a_{\alpha,\tau}^{\dagger}$, $a_{\alpha,\tau}$ as $a_{\alpha,\tau}\ket{O_{\tau}}=0$. The two sets of operators at different times are related by a time-dependent unitary Bogoliubov transformation $a_{\alpha,\tau} = \mathcal{W}_{\tau}^{\dagger}\, b_{\alpha}\, \mathcal{W}_{\tau}$, where the unitary operator $\mathcal{W}_{\tau}$ has the property to relate the different vacuum states through $\ket{O_{\tau_-}} =\mathcal{W}_{\tau} \ket{O_{\tau}}$. Combining this expression with Eq. (\ref{equ:solution_with_operator_evolution}), we observe that the state of $\phi$ at any given instant can be expressed from the vacuum at this moment according to
\begin{equation}
 \label{equ:final_state_related_to_vacuum}
\ket{\phi(\tau)}= \mathcal{U}_{\tau} \mathcal{W}_{\tau} \ket{O_{\tau}}:=\mathcal{X}_{\tau} \ket{O_{\tau}}.
\end{equation}
Starting from the vacuum, the field has evolved to an excited state whose spectrum is encompassed in the nonunitary operator $\mathcal{X}_{\tau}$. The main result of Rubakov was to explicitly write this operator in terms of the creation-annihilation operators $a_{\alpha,\tau}^{\dagger}$, $a_{\alpha,\tau}$, thus acting directly on their related vacuum state $\ket{O_{\tau}}$. He used the method of nonunitary (resp. unitary) Bogoliubov transformations to compute $\mathcal{U}_{\tau}$ (resp. $\mathcal{W}_{\tau}$). We do not give the details of this calculation and directly present the prescription to find $\mathcal{X}_{\tau}$. 

At any instant $\tau$, the state of $\phi$ is given by\footnote{We stress that this formula is only valid if  $\ket{\phi(\tau_{-})}=\ket{O_{\tau_-}}$. Rubakov [\cite{Rubakov:1984}, Eq. (3.16)] gave the more complete expression for an initially excited state.}
\begin{equation}
\label{equ:phi_state_evolved}
\ket{\phi(\tau)}=\mathcal{X}_{\tau}\ket{O_{\tau}}= C(\tau) \exp \left( \frac12 \sum_{\alpha, \beta} D_{\alpha \beta}(\tau) a_{\alpha,\tau}^{\dagger} a_{\beta,\tau}^{\dagger} \right) \ket{O_{\tau}}
\end{equation}
and the number of particles in a given mode $\alpha$ is then
\begin{equation}
\label{equ:number_created_particle}
N_{\alpha}(\tau)= \frac{\braket{\phi(\tau)|a_{\alpha,\tau}^{\dagger}a_{\alpha,\tau}|\phi(\tau)}}{\braket{\phi(\tau)|\phi(\tau)}}= \left( \frac{D^2(\tau)}{\mathbbm{1}-D^2(\tau)}\right)_{\alpha \alpha},
\end{equation}
where $C(\tau)$ is a $\tau$ dependent c-number\footnote{This prefactor $C$ cannot be computed from the Bogoliubov transformations, but as we will discuss later we do not need it to extract the backreaction of particle production.} and $D$ is a real symmetric matrix defined as follows. The two key ingredients to compute $D$ are the sets of functions $\left\{\xi_{\tau}^{\alpha}(\mathbf{x}) \right\}$ and $\left\{h_{\alpha}(\mathbf{x}, \tau) \right\}$. At each instant $\tau$, the $\xi_{\tau}^{\alpha}$ form the complete set of real wave functions of the $\phi$ particles in the background of the bounce. So, they satisfy
\begin{equation}
 \label{equ:xi_differential_equations}
\left[-\partial_i^2 +m^2(\mathbf{x},\tau) \right]\xi_{\tau}^{\alpha}(\mathbf{x}) =\left(\omega_{\tau}^{\alpha} \right)^2\xi_{\tau}^{\alpha}(\mathbf{x})
\end{equation}
with the normalization condition
\begin{equation}
 \label{equ:xi_normalization}
\left( \xi_{\tau}^{\alpha}, \xi_{\tau}^{\beta}\right):= \int d^3 \mathbf{x} \ \xi_{\tau}^{\alpha}(\mathbf{x}) \, \xi_{\tau}^{\beta}(\mathbf{x})= \left( 2 \omega_{\tau}^{\alpha}\right)^{-1} \delta_{\alpha \beta}.
\end{equation}
On the other hand, the $h_{\alpha}$ form the set of real Euclidean mode functions similar to the positive-frequency functions in Minkowski spacetime. So, they satisfy the Euclidean field equation
\begin{equation}
 \label{equ:mode_function_differential_equations}
\left[ - \partial_{\tau}^2-\partial_i^2 +m^2(\mathbf{x},\tau)\right]h_{\alpha}(\mathbf{x},\tau)=0
\end{equation}
with the condition that they exponentially decrease to zero for $\tau \to - \infty$. 
The last step is to construct the above matrix as
\begin{equation}
\label{equ:matrix_D_V_Z}
D=V Z^{-1},
\end{equation}
where
\begin{equation}
\label{equ:matrices_V_Z}
\begin{split}
V_{\alpha \beta}(\tau)=\left( \omega_{\tau}^{\alpha} \xi_{\tau}^{\alpha}, h_{\beta} \right) - \left( \xi_{\tau}^{\alpha}, \partial_{\tau} h_{\beta}\right) \\
Z_{\alpha \beta}(\tau)=\left( \omega_{\tau}^{\alpha} \xi_{\tau}^{\alpha}, h_{\beta} \right) + \left( \xi_{\tau}^{\alpha}, \partial_{\tau} h_{\beta}\right)
\end{split}.
\end{equation}

In summary, the spectrum of created particles described by Eqs. (\ref{equ:phi_state_evolved}) and (\ref{equ:number_created_particle}) is entirely defined by the solutions of the differential equations (\ref{equ:xi_differential_equations}) and (\ref{equ:mode_function_differential_equations}). However, the computation of the matrix $D$ can be a very difficult task since it requires us to invert a generally infinite-dimensional matrix. In the case of a homogeneous decay, we shall see in Sec. \ref{sec:toy_model} that this computation can be performed straightforwardly because the above matrices are diagonal. However, this becomes often impossible in the case of bubble nucleation. In order to avoid this problem, Rubakov actually provided an iterative way to estimate $D$. Before presenting this method in the next section, it is worth making some remarks on the above result. 

First, the number of created particles (\ref{equ:number_created_particle}) has only a physical meaning at $\tau=0$ when the bubble materializes. Second, we observe that this result diverges if the matrix $\mathbbm{1}-D^2$ has a zero eigenvalue. According to Rubakov, this problem could occur for the self-excitation of the tunneling field $\sigma$ itself. So we should remember this remark if we want to consider such a case. However, it is important to realize that this divergence is not related to the common ultraviolet divergences of quantum field theories. Actually, Rubakov proved that the total number of created particles per bubble $N = \sum_{\alpha} N_{\alpha}$ is ultraviolet finite. As this result will be important for our own discussion regarding the backreaction of these particles, we shall discuss the reason of this fact below. 


\subsection{\label{sub:weak_particle}Weak particle production and UV finiteness}
Let us now present how the matrix $D$ can be approximated without inverting an infinite-dimensional matrix. It relies on the observation that $D$ satisfies the following matrix differential equation:
\begin{equation}
\label{equ:iterative_equation_D_matrix}
\partial_{\tau}D= -(ED+DE)+B+(AD-DA)-DBD
\end{equation}
where $E$, $A$ and $B$ are some matrices given by Rubakov (\cite{Rubakov:1984}, Sec. 3.3.3):

\begin{equation}
\label{equ:matrices_iteration}
E_{\alpha \beta}=\omega_{\tau}^{\alpha} \delta_{\alpha \beta}, \, A_{\alpha \beta }= \left\{\begin{array}{c}
 0, \, \alpha = \beta \\
\frac1{\omega_{\tau}^{\alpha} - \omega_{\tau}^{\beta}} \left( \xi_{\tau}^{\alpha}, \partial_{\tau}m^2 \xi_{\tau}^{\beta}\right), \,  \mbox{}_{\alpha \neq \beta} \\
\end{array},\right.  B_{\alpha \beta }=\frac1{\omega_{\tau}^{\alpha} + \omega_{\tau}^{\beta}} \left( \xi_{\tau}^{\alpha}, \partial_{\tau}m^2 \xi_{\tau}^{\beta}\right).
\end{equation}
We note that $D$ is now uniquely specified by the wave functions $\xi_{\tau}^{\alpha}$, without reference to $h_{\alpha}$. Moreover, when the parameter $\partial_{\tau} m^2$ is small [typically compared to $(\omega_{\tau}^{\alpha})^3$], the matrices $A$ and $B$ can be considered as a perturbation and Eq. (\ref{equ:iterative_equation_D_matrix}) can be solved iteratively. The lowest order solution of Eq. (\ref{equ:iterative_equation_D_matrix}) is given by the formula (3.28) in \cite{Rubakov:1984}:
\begin{equation}
\label{equ:matrix_D_weak_limit}
D_{\alpha \beta}(\tau)= e^{-\left(W_{\alpha}(\tau)+W_{\beta}(\tau)\right)} \int_{-\infty}^{\tau} e^{\left(W_{\alpha}(\tau')+W_{\beta}(\tau')\right)} \frac{\left( \xi_{\tau'}^{\alpha}, \partial_{\tau'} m^2 \xi_{\tau'}^{\beta}\right)}{\omega_{\tau'}^{\alpha}+\omega_{\tau'}^{\beta}} d\tau',
\end{equation}
where $W_{\alpha}(\tau)=\int_{0}^{\tau} d\tau' \omega_{\tau'}^{\alpha}$.
This limit corresponds to say that the number of created particles is weak and the matrix $D$ is small. Thus, Eq. (\ref{equ:number_created_particle}) for the number of created particles in a given mode and the total number of particles reduce to
\begin{equation}
\label{equ:number_created_particle_weak}
N_{\alpha}(\tau)=  \left( D^2(\tau)\right)_{\alpha \alpha} \quad \text{and} \quad N=\Tr D^2.
\end{equation}

The two previous equations are also useful to prove that the total number of created particles is ultraviolet finite. Indeed, $\partial_{\tau} m^2$ becomes clearly negligible in front of $(\omega_{\tau}^{\alpha})^3$ for high energy particles. Moreover, the $m^2$ term can be neglected in Eq. (\ref{equ:xi_differential_equations}), such that the wave functions $\xi_{\tau}^{\alpha}$ simply become in that sector\footnote{Although the formalism was built with real wave functions $\xi_{\tau}^{\alpha}$, there is no difficulty to consider complex functions in this short discussion.}
\begin{equation}
\label{equ:wave_function_high_energy}
\xi_{\tau}^{\alpha}(\mathbf{x}) = \frac{e^{i \mathbf{k} \mathbf{x}}}{(2\pi)^{3/2} \sqrt{2 \omega_{\tau}^{\mathbf{k}}}} \quad \text{ with } \quad  \omega_{\tau}^{\mathbf{k}} = k=|\mathbf{k}|.
\end{equation}
Plugging Eq. (\ref{equ:wave_function_high_energy}) in Eqs. (\ref{equ:matrix_D_weak_limit}) and (\ref{equ:number_created_particle_weak}), Rubakov \cite{Rubakov:1984} concluded that the number of particles in the UV region roughly behaves as
 \begin{equation}
\label{equ:number_high_energy_particles}
N_{UV}  = \int d^3 \mathbf{k} d^3 \mathbf{k'} |D_{\mathbf{k} \mathbf{k'}}|^2  \propto \int d^3 \mathbf{k} d^3 \mathbf{k'} \frac{1}{ k k' (k+k')^4} \left|\widetilde{\partial_{\tau} m^2}(\mathbf{k}-\mathbf{k}', \tau)  \right|^2,
\end{equation}
where $\widetilde{\partial_{\tau}m^2}$ is the Fourier transform of $\partial_{\tau}m^2$. The above integral converges if $\widetilde{\partial_{\tau}m^2}$ rapidly goes to zero when $|\mathbf{k}-\mathbf{k}'| \rightarrow +\infty$. This is actually the case since at high momentum $m \ll \omega_{\tau}^{\mathbf{k}} \approx k$. It means that in the ultraviolet sector the system is insensitive to the variations of the background $\partial_{\tau}m^2$ and this naturally regularizes the expression (\ref{equ:number_high_energy_particles}).


\section{\label{sec:backreaction}Backreaction of particle production}
\subsection{Reduced density matrix formalism}
In the previous section, we assumed that the external field $\phi$ had no impact on the tunneling process. We propose now an approach to estimate the effect of the created particles on the semiclassical decay rate of $\sigma$. The idea is to work with the reduced density matrix which is a tool introduced in the early days of quantum mechanics by Landau \cite{Landau:1927}. Its use is convenient to investigate the impact of the environment on a given quantum system. Let us briefly expose how this mechanism works at the general level.

Consider a system $S$ of interest described by a Hilbert space $\mathcal{H}_S$ coupled to an environment $\mathcal{E}$ with the Hilbert space $\mathcal{H_E}$. The total state vector satisfies $\ket{\Psi}\in \mathcal{H}_S \otimes \mathcal{H_E}$ and the total density operator is then given by
\begin{equation}
\label{equ:density_matrix_general}
\hat{\rho}= \ket{\Psi}\bra{\Psi}.
\end{equation}
We suppose that we are interested in an observable $\hat{A}$ which is only related to the system $S$ and not to the environment $\mathcal{E}$. In other words, we can write it as $\hat{A}=\hat{A}_S \otimes \hat{I}_{\mathcal{E}}$ where $\hat{A}_S$ acts on $\mathcal{H}_S$ and $I_{\mathcal{E}}$ is the identity acting on the environment. An important consequence from quantum mechanics is that the measurement of $\hat{A}$ satisfies \cite{Zurek:2003,Schlosshauer:2003} 
\begin{equation}
\label{equ:measurement_matrix_and_reduced}
\langle \hat{A} \rangle_{\Psi}= \Tr \left( \hat{\rho} \hat{A} \right) = \Tr_{\mathcal{H}_S} \left( \hat{\rho}_S \hat{A}_S \right),
\end{equation}
where $\hat{\rho}_S$ is the reduced density operator obtained by tracing over the environment:
\begin{equation}
\label{equ:reduced_densty_matrix_quantum}
\hat{\rho}_S = \Tr_{\mathcal{H_E}}\left(\hat{\rho}\right).
\end{equation}
The right-hand side of Eq. (\ref{equ:measurement_matrix_and_reduced}) tells us that the effect of the environment only enters in the reduced density operator and thus it entirely contains the effect of these external degrees of freedom.

We can now apply this formalism to our situation as well. Indeed, the role of $S$ and $\mathcal{E}$ is played by $\sigma$ and $\phi$ respectively. Moreover, since the decay rate is a quantity which is only related to $\sigma$, the backreaction of the created $\phi$ particles is entirely encompassed in the reduced density operator. We shall now explicitly compute it and discuss its impact on $\Gamma$. From the expressions (\ref{equ:total_state_separation_bounce}) and (\ref{equ:phi_state_evolved}) describing the state vector of the model of the two scalar fields, we can write the density operator as
\begin{equation}
\label{equ:density_matrix_X_operator}
\hat{\rho}(\tau)= \ket{\Psi(\tau)}\bra{\Psi(\tau)} = e^{-2 S_E[\sigma_B(\tau)]} \left( \mathcal{X}_{\tau} \ket{O_{\tau}}\bra{O_{\tau}} \mathcal{X}^{\dagger}_{\tau}\right).
\end{equation}
In this context, tracing over the environment means to sum over all possible particle states of the field $\sigma$. So we define 
\begin{equation}
\label{equ:general_state_number_boson}
\ket{ \left\{\alpha\right\}_n(\tau)}= \prod_{i=1}^{n} a_{\alpha_i,\tau}^{\dagger}  \ket{O_{\tau}},
\end{equation}
which describes an unnormalized $n$-particle state with each particle in a given mode $\alpha_i$: $\left\{\alpha\right\}_n=\left\{\alpha_1,\alpha_2,\ldots,\alpha_n\right\}$. As it should be the case for bosons, no restriction on $\left\{\alpha\right\}_n$ is imposed, meaning that two or more particles can be in the same state, as for example if $\alpha_i=\alpha_j$ for some $i\neq j$. To obtain the reduced density operator, we first apply the state (\ref{equ:general_state_number_boson}) on both sides of (\ref{equ:density_matrix_X_operator}) and then we have to sum over all the possible configurations $\left\{\alpha\right\}_n$ and all the numbers of particles. Thus,
\begin{equation}
\label{equ:reduced_density_matrix_X_operator}
\hat{\rho}_r(\tau)= e^{-2 S_E[\sigma_B(\tau)]} \sum_{n=0}^{\infty} \sum_{\left\{ \alpha \right\}_n} \frac{\left| \braket{ \left\{ \alpha \right\}_n(\tau)|\mathcal{X}_{\tau}| O_{\tau}}\right|^2}{n!},
\end{equation}
where we have introduced the factor $n!$ to ensure the correct normalization. In this formula, the dependence on the modes appears as a sum and can easily be extended to an integral in the case of a continuous index $\alpha$. For convenience, we shall keep the summation symbol throughout this section.

The decay rate of $\sigma$ including the backreaction of the $\phi$ particles is now given by
\begin{equation}
\label{equ:decay_probability_with_backreaction}
\frac{\Gamma_{\text{Ba}}}{\mathcal{V}} = \frac{\hat{\rho}_r(0)}{\hat{\rho}_r(- \infty)} = e^{-S_B} \frac{\mathcal{F}(0)}{\mathcal{F}(-\infty)} \left|C(0) \right|^2,
\end{equation}
where we have used $S_B=2 S_E[\sigma_B(\tau=0)]$ and $\mathcal{F}(\tau)$ is defined as
\begin{equation}
\label{equ:F_curl_factor}
\mathcal{F}(\tau) = \sum_{n=0}^{\infty} \sum_{\left\{ \alpha\right\}_n} \frac{\left| \braket{ \left\{ \alpha \right\}_n(\tau)|\exp{\left(\frac12 \sum_{\alpha,\beta} D_{\alpha\beta}(\tau) a^{\dagger}_{\alpha,\tau} a^{\dagger}_{\beta,\tau}\right)}| O_{\tau}}\right|^2}{n!}.
\end{equation}
As expected, we recover in Eq. (\ref{equ:decay_probability_with_backreaction}) the semiclassical exponential and the new correction factor. The main purpose of the next section is to find a convenient expression for $\mathcal{F}(\tau)$. 


\subsection{\label{sub:explicit_computation}Explicit computation}
Since we shall work at a given instant, we omit the $\tau$ dependence in the expression (\ref{equ:F_curl_factor}) for $\mathcal{F}$. We first simplify the matrix element $M_{\left\{ \alpha\right\}_n}:=\braket{ O|a_{\alpha_1}\ldots a_{\alpha_{n}}\exp{\left(\frac12 \sum_{\alpha,\beta} D_{\alpha\beta}a^{\dagger}_{\alpha}a^{\dagger}_{\beta}\right)}| O}$. Once we expand the exponential as the usual power series, we observe that the only nonvanishing contribution comes from the term which contains the same number of creation operators $a^{\dagger}$ as the number of annihilation operators $a$ on their left. It implies in particular that the matrix element vanishes for $n$ odd. So with $n=2k$, we get
\begin{equation}
 \label{equ:matrix_element_exponential_simplified}
\begin{split}
& M_{\left\{ \alpha \right\}_n}  = M_{\left\{ \alpha \right\}_{2k}}  =\frac{1}{2^k k!}\braket{O| a_{\alpha_1}\ldots a_{\alpha_{2k}} \left(\sum_{\alpha,\beta} D_{\alpha\beta}a^{\dagger}_{\alpha}a^{\dagger}_{\beta}\right)^k| O} \\
& = \frac{1}{2^k k!} \sum_{\beta_1, \ldots, \beta_{2k}} D_{\beta_1\beta_2}\ldots D_{\beta_{2k-1}\beta_{2k}} \braket{O| a_{\alpha_1}\ldots a_{\alpha_{2k}} a^{\dagger}_{\beta_1}\ldots a^{\dagger}_{\beta_{2k}}|O},
\end{split}
\end{equation}
where we have factorized the $D_{\alpha \beta}$ since they are real numbers. The remaining matrix element can be computed straightforwardly from the commutation rules of the bosonic operators ($[a_{\alpha},a^{\dagger}_{\beta}]=\delta_{\alpha \beta}$ and zero otherwise):
\begin{equation}
\label{equ:commutation_boson_matrix_element}
\braket{O|a_{\alpha_1}\ldots a_{\alpha_{2k}}a^{\dagger}_{\beta_{1}}\ldots a^{\dagger}_{\beta_{2k}}|O}=\braket{O|O} \sum_{\pi \in S_{2k}} \delta_{ \beta_1 \alpha_{\pi(1)}} \ldots \delta_{\beta_{2k} \alpha_{\pi(2k)}}
\end{equation}
where $\pi$ is a permutation of the symmetric group $S_{2k}$. Once (\ref{equ:commutation_boson_matrix_element}) is introduced into (\ref{equ:matrix_element_exponential_simplified}), each delta symbol selects one term of each sum over the modes $\beta_i$ and thus
\begin{equation}
 \label{equ:matrix_element_delta_simplified}
M_{\left\{ \alpha \right\}_{2k}}  = \braket{O|O} \frac{1}{2^k k!} \sum_{\pi \in S_{2k}} \prod_{i}^k D_{\alpha_{\pi(2i-1)}\alpha_{\pi(2i)}}.
\end{equation}

The initial expression (\ref{equ:F_curl_factor}) actually involves $\left| M_{\left\{ \alpha \right\}_{2k}}\right|^2$. As $D$ is a real matrix, it reduces to take the square of Eq. (\ref{equ:matrix_element_delta_simplified}). As usual, the square of a sum can be written as a double sum and we find
\begin{equation}
\label{equ:F_curl_simplified_intermediate}
\mathcal{F} = \left|\braket{O|O}\right|^2 \sum_{k=0}^{+\infty} \frac1{(2k)! (k!)^2 2^{2k}} \sum_{\left\{ \alpha \right\}_{2k}} \sum_{\pi, \sigma \in S_{2k}} \prod_{i=1}^k D_{\alpha_{\pi(2i-1)}\alpha_{\pi(2i)}} D_{\alpha_{\sigma(2i-1)}\alpha_{\sigma(2i)}}.
\end{equation}
Thanks to the summation over $\left\{ \alpha\right\}_{2k}$, we can relabel the dummy indices in the previous equation as $\alpha_{\pi(i)} \rightarrow \alpha_i$ and  $\alpha_{\sigma(i)} \rightarrow \alpha_{\pi^{-1}(\sigma(i))}$. It allows us to factorize the product $\prod_{i=1}^k D_{\alpha_{2i-1}\alpha_{2i}}$ in front of the summation over $\pi$ and $\sigma$ and then to use the identity
\begin{equation}
\label{equ:double_permutation_summation}
\begin{split}
\sum_{\pi, \sigma \in S_{2k}} \prod_{i=1}^k D_{\alpha_{\pi^{-1}(\sigma(2i-1))} \alpha_{\pi^{-1}(\sigma(2i))}} = (2k)! \sum_{\lambda \in S_{2k}} \prod_{i=1}^k D_{\alpha_{\lambda(2i-1)} \alpha_{\lambda(2i)}}. \\
\end{split}
\end{equation}
This equality holds because for each one of the $(2k)!$ permutations $\pi$, the sum over $\sigma$ is always the same up to the order of the terms. The reason behind this statement comes from the group property of $S_{2k}$: when $\pi$ is fixed and $\sigma$ runs over all the elements of the symmetric group, $\pi^{-1}\circ \sigma$ covers the whole group as well. So, we have reduced the expression to a single sum over the permutations and moreover the two factors $(2k)!$ cancel out:
\begin{equation}
 \label{equ:F_curl_A_k}
\mathcal{F}=\left|\braket{O|O}\right|^2 \sum_{k=0}^{\infty} \underbrace{ \frac{1}{(k!)^2 2^{2k} } \sum_{\left\{ \alpha \right\}_{2k}} \sum_{\lambda \in S_{2k}} \prod_{i=1}^k D_{\alpha_{2i-1}\alpha_{2i}} D_{\alpha_{\lambda(2i-1)}\alpha_{\lambda(2i)}}}_{:=A_k}.
\end{equation}

To understand how to further simplify this expression, it is worth writing the first terms $A_k$ explicitly. Since the matrix $D$ is symmetric, the summation over the modes $\alpha_1, \alpha_2,\ldots$ gives the trace of an even power of the matrix $D$, as for instance
\begin{equation}
\sum_{\alpha_1,\alpha_2} D_{\alpha_1 \alpha_2} D_{\alpha_1 \alpha_2}=\sum_{\alpha_1,\alpha_2} D_{\alpha_1 \alpha_2} D_{\alpha_2 \alpha_1}=\sum_{\alpha_1} \left(D^2\right)_{\alpha_1 \alpha_1} =\Tr\left( D^2\right).
\end{equation}
So a straightforward computation gives us the first four terms:
\begin{equation}
 \label{equ:F_curl_first_k}
\begin{array}{ccl}
 A_0  & = & 1 \\
 A_1  & = & \frac12 \Tr\left( D^2\right)  \\
 A_2  & = & \frac18 \Tr\left( D^2\right)^2 + \frac14  \Tr\left( D^4\right) \\
 A_3  & = & \frac{1}{48} \Tr\left( D^2\right)^3 + \frac18  \Tr\left( D^2\right) \Tr\left( D^4\right) + \frac16 \Tr\left( D^6\right).
\end{array}
\end{equation}
It is clear from both (\ref{equ:F_curl_A_k}) and (\ref{equ:F_curl_first_k}) that each term $A_k$ is a sum of products of the following form: $\Tr(D^2)^{j_1} \Tr(D^4)^{j_2}\ldots \Tr(D^{2k})^{j_k}$, where the integers $j_i$ satisfy $j_1+2j_2+\ldots+ k j_k = k$. Such a set $\left\{ j_i\right\}_k$ is called a partition of $k$. Thus, $A_k$ can be written as a sum over all the partitions of $k$ as
\begin{equation}
\label{equ:term_A_k_unknown_coeff}
A_k = \sum_{\left\{ j_i\right\}_k} \frac{C_{\left\{ j_i\right\}_k}}{(k!)^2 2^{2k}} \prod_{i=1}^{k} \Tr\left(D^{2i}\right)^{j_i},
\end{equation}
where each coefficient $C_{\left\{ j_i\right\}_k}$ counts the number of permutations $\lambda \in S_{2k}$ in (\ref{equ:F_curl_A_k}) which lead to the partition $\left\{ j_i\right\}_k$. In particular, they satisfy $\sum_{\left\{ j_i\right\}_k} C_{\left\{ j_i\right\}_k}=(2k)!$. According to a combinatorial argument given in Appendix \ref{app:combinatorial}, we find an explicit formula for them:
\begin{equation}
\label{equ:ch4_coefficient_C_k_combinatoric}
C_{\left\{ j_i\right\}_k}= \frac{(k!)^2 2^{2k}}{\prod_{i=1}^{k} j_i! (2i)^{j_i}}.
\end{equation}
This result significantly simplifies $A_k$ and $\mathcal{F}$. Indeed, if we insert it in (\ref{equ:term_A_k_unknown_coeff}) and define for convenience the variables $a_i:=\Tr\left(D^{2i}\right) i! / (2i)$, we find that $A_k$ actually corresponds to a complete Bell polynomial $B_k(a_1,\ldots,a_k)$ \cite{Bell:1927}:
\begin{equation}
\label{equ:ch4_term_A_k_d_rond_with_coeff}
A_k  = \sum_{\left\{ j_i\right\}_k} \prod_{i=1}^{k} \frac1{j_i!}\left(\frac{a_i}{i!}\right)^{j_i}  = \frac{1}{k!} B_k( a_1,a_2,\ldots,a_k).
\end{equation}
It is a well-known fact in combinatorics that these polynomials give an exponential formula for	any formal series $a_1 x +\cdots + \frac{a_n}{n!}x^n +\cdots$ \cite{Comtet:1974}:
\begin{equation}
\label{equ:ch4_expoential_bell_formal_series}
\sum_{k=0}^{+\infty}\frac{B_k(a_1,\ldots, a_k)}{k!}x^k=\exp\left( \sum_{k=1}^{+\infty} \frac{a_k}{k!}x^k \right).
\end{equation}
If we set $x=1$, the left-hand side of this equation is exactly the sum over the terms $A_k$. So it is now straightforward to conclude that $\mathcal{F}$ is given by the following expressions
\begin{equation}
\label{equ:F_curl_explicit}
\begin{split}
 \mathcal{F} & = \left|\braket{O|O}\right|^2  \exp \left( \sum_{k=1}^{\infty} \frac{\Tr \left(D^{2k} \right)}{2k}  \right) \\
  & = \left|\braket{O|O}\right|^2  \exp \left( -\frac12 \Tr \ln \left( \mathbbm{1}-D^2\right) \right) \\
  & = \left|\braket{O|O}\right|^2 \frac1{\sqrt{\Det \left(\mathbbm{1}- D^2\right)}}.
\end{split}
\end{equation}

From this result, we can now derive a convenient expression for the backreaction factor $\frac{\mathcal{F}(0)}{\mathcal{F}(-\infty)}$  in Eq. (\ref{equ:decay_probability_with_backreaction}). Since $D(\tau=-\infty)=0$ and $\braket{O_{\tau=-\infty}|O_{\tau=-\infty}}=\braket{O_{\tau}|\mathcal{W}_{\tau}^{\dagger}\mathcal{W}_{\tau}|O_{\tau}}=\braket{O_{\tau}|O_{\tau}}$, we obtain the modified decay rate as follows:

\begin{equation}
\label{equ:decay_probability_backreaction_final}
\frac{\Gamma_{\text{Ba}}}{\mathcal{V}} = \frac{ \left|C_{\tau=0} \right|^2 e^{-S_B}}{\sqrt{\Det \left(\mathbbm{1}- D^2_{\tau=0}\right)}} = \left|C_{\tau=0}\right|^2 e^{-S_B - \frac12 \Tr \ln \left( \mathbbm{1}-D^2_{\tau=0}\right) } .
\end{equation}

In this expression, the backreaction of the real $\phi$ particles produced during the decay of the field $\sigma$ is encompassed in the prefactor $\left[\Det \left(\mathbbm{1}- D^2_{\tau=0}\right)\right]^{-1/2}$. On the other hand, the coefficient $\left|C_{\tau=0} \right|^2$ can be evaluated from the well-known path integral method of Callan and Coleman \cite{Callan:1977}, corresponding to one-loop quantum fluctuations of $\phi$. We shall now exclusively focus our attention on the properties of the new backreaction factor.


\subsection{\label{sub:interpretation}Interpretation of the backreaction factor}

Without considering any particular model, Eq. (\ref{equ:decay_probability_backreaction_final}) gives interesting information on the effect of particle production during tunneling. The first remark is to realize that our derivation has been formal and that we should pay attention to potential divergences. As it was already the case for the number of particles (\ref{equ:number_created_particle}), we observe again a problematic behavior when $\mathbbm{1}-D^2$ has a zero eigenvalue. As already explained, this has nothing to do with the usual UV divergence and this problem should be treated on a case-by-case basis. More interestingly, we can show that the backreaction factor is UV finite. The reason is clear if we look at the first line of Eq. (\ref{equ:F_curl_explicit}). Indeed, we already showed in Eq. (\ref{equ:number_high_energy_particles}) that the first term $\Tr D^2$ is finite. It is then straightforward to extrapolate the argument to each following term, $\Tr  \left( D^{2k} \right)$ for $k=2,3,\ldots$, since they will converge faster and faster. We conclude, in contrast with one-loop quantum corrections, that the backreaction factor does not require any renormalization technique in order to be computed. 

The second remark is related to the weak particle production limit. As we explained in Sec. \ref{sub:weak_particle}, the total number of created particles is given in this approximation by $N \approx \Tr D^2$ since $D$ is small. For the same reason, the terms $\Tr  \left( D^{2k} \right)$  with $k>2$ appearing in Eqs. (\ref{equ:F_curl_explicit}) and (\ref{equ:decay_probability_backreaction_final}) are negligible in front of the dominant term $\Tr D^2$. It means that in this limit, the modified decay rate becomes $\frac{\Gamma_{\text{Ba}}}{V} \approx e^{-S_B + \frac12 N }$. In other words, we see that the backreaction is directly given by one half of the total number of produced particles. This is a useful result if we are interested in an approximated value of the backreaction, since its evaluation does not require us to perform any computation in addition to Rubakov's prescription to find $N$.

The third observation of interest is that the backreaction of scalar particle production enhances the semiclassical decay probability in this given framework. Looking at Eq. (\ref{equ:decay_probability_backreaction_final}) or at the approximate formula we just discussed, $\frac{\Gamma_{\text{Ba}}}{V} \approx e^{-S_B + \frac12 N }$, we observe that the backreaction contribution is positive compared to the semiclassical ($-S_B)$ contribution. At first sight, this result might seem surprising. During tunneling, the field $\sigma$ transfers some amount of energy to the environment $\phi$ which consequently exhibits a spectrum of particles. We may have expected that this dissipation from the system to the surrounding bath slows down the decay process. However, the created particles are fluctuations of the field $\phi$ which in turn reacts on $\sigma$. We are actually facing a situation which has some similarities with the fluctuation-dissipation relation in statistical physics \cite{Callen:1951}. Both the system and the environment interact with each other in a nontrivial way and eventually Eq. (\ref{equ:decay_probability_backreaction_final}) tells us that the tunneling process is enhanced in this particular case. This kind of behavior has already been discussed in a variety of situations. The investigation of the impact of the environment on the tunneling of a quantum mechanical particle was initiated by Caldeira and Leggett in \cite{Caldeira:1983}. By considering a dissipative interaction with a bath of harmonic oscillators, they derived a friction term suppressing the decay rate. On the other hand, \cite{Bruinsma:1986,Fujikawa:1992} subsequently described systems for which the impact of the environment results in an enhancement of the tunneling probability of the quantum particle. More closely related to our case, \cite{Calzetta:2001,Kiefer:2010} also reported an enhanced decay rate for a quantum field interacting with some external degrees of freedom.

Our next objective is to quantitatively estimate the contribution of the backreaction in comparison to the semiclassical decay exponent $S_B$. This requires us to explicitly evaluate the formula (\ref{equ:decay_probability_backreaction_final}) and thus to consider some specific models.


\section{\label{sec:toy_model}Toy model potential}
In order to illustrate the mechanism of particle production and the related backreaction, we consider the following tractable potential:
 \begin{equation}
 \label{equ:toy_potential}
 V(\sigma) = \frac{\lambda}{8} \left(\sigma^2-v^2\right)^2-\frac{\epsilon}{2v} (\sigma+v),
 \end{equation}
where  $\lambda$, $v$ and $\epsilon$ are three parameters with dimension $[\lambda]=E^0$, $[v]=E^1$ and $[\epsilon]=E^4$. This model is widely used in the literature since the bounce solution can be computed analytically in the limit of small $\epsilon$. In particular, Coleman \cite{Coleman:1977} introduced it to show how to compute the semiclassical decay rate $S_B$ and Rubakov \cite{Rubakov:1984} used it to illustrate its prescription of particle production. Our aim is now to extend these investigations by estimating the backreaction of the created particles.

As already discussed in Sec. \ref{sub:decay_rate}, we can focus on two different types of decay, namely the nucleation of an $O(4)$ bubble or a homogeneous process in a finite volume $\mathcal{V}$. We shall consider both of these cases in the approximation that the energy difference between the two vacua of the potential $V(\sigma)$ is much smaller than the height of the barrier. It is not difficult to show that it corresponds to a small value of $\epsilon$. Indeed, in that case, the vacua are approximately located at $\sigma_{T/F}\approx \pm v$, the energy difference becomes $\Delta V=|V(\sigma_F)-V(\sigma_T)|\approx \epsilon$ and the barrier height is given by $V_B\approx V(0)\approx \frac18 \lambda v^4$.  As expected, the requirement $\Delta V \ll V_B$ is fulfilled for 
 \begin{equation}
 \label{equ:condition_small_epsilon}
 \epsilon \ll \frac18 \lambda v^4.
 \end{equation}

We now compute the semiclassical decay exponent $S_B$ in the two cases of interest:

\begin{enumerate}
\item Thin-wall (TW) bounce

Let us focus first on the bubble nucleation under the condition (\ref{equ:condition_small_epsilon}). The bounce satisfying Eq. (\ref{equ:bounce_equation_o4}) is well approximated by the kink
 \begin{equation}
 \label{equ:solution_field_equation_thin_wall}
 \sigma(\rho ) = v \tanh \left( \frac12 v \sqrt{ \lambda} (R-\rho)\right),
 \end{equation}
 where $R$ is the radius of the bubble at the nucleation time $\tau=0$. The value of this parameter which minimizes the action (\ref{equ:euclidean_action_o4_bubble}) is given by
\begin{equation}
 \label{equ:R_thin_wall_bounce}
 R^{\star}=\frac{2 \sqrt{\lambda} v^3 }{\epsilon}.
\end{equation}
The semiclassical decay exponent is then computed by evaluating the action (\ref{equ:euclidean_action_o4_bubble}) along the solution (\ref{equ:solution_field_equation_thin_wall}) with this value of $R^{\star}$. This gives
 \begin{equation}
 \label{equ:decay_rate_exponent_thin_wall_bounce}
 S_{B,TW} = \frac83 \pi^2 \frac{\lambda^2 v^{12} }{\epsilon^3}.
 \end{equation}
 We now understand why this solution is called a thin-wall bubble. Under the approximation of small $\epsilon$, its radius (\ref{equ:R_thin_wall_bounce}) is large while the transition wall between the false and true vacua is thin because of the $\tanh$ shape of Eq. (\ref{equ:solution_field_equation_thin_wall}). 
 
\item Homogeneous bounce

We consider a spatially homogeneous solution $\sigma_B(\tau)$ of Eq. (\ref{equ:bounce_equation_general}) in a sphere of radius $R_{H}$. For a small $\epsilon$, the result is again the kink: \begin{equation}
\label{equ:solution_field_equation_toy_homogeneous}
\sigma(\tau ) = v \tanh \left(\frac12 v \sqrt{ \lambda} \tau  +C \right).
\end{equation}
The constant $C$ is chosen in order to satisfy the condition that the field emerges under the barrier at $\tau=0$. Since the escape point is given by $\sigma_{esc} = v -\frac{1}{v} \sqrt{\frac{2\epsilon}{\lambda }}+ O(\epsilon)$, we have 
\begin{equation}
\label{equ:turning_point_and_C_toy_homogeneous}
 \tanh(C) \approx 1 - \frac{1}{v^2}\sqrt{\frac{2 \epsilon}{\lambda}}.
 \end{equation}
We can evaluate the Euclidean action (\ref{equ:euclidean_action_homogeneous}) along the bounce taking into account the above value of $C$ and the volume $\mathcal{V}=\frac43 \pi R_H^3$.\footnote{Note that the solution (\ref{equ:solution_field_equation_toy_homogeneous}) only covers one half of the bounce ($\tau<0$) and that $\sigma(-\tau)$ gives the $\tau>0$ part.} In the limit of vanishing $\epsilon$, we get
\begin{equation}
\label{equ:SB_coefficient_homogeneous}
S_{B,hom} =  \frac{16}{9} \pi R_H^3 \sqrt{\lambda} v^3.
\end{equation}
Looking at the $\tanh$ form of the bounce, we observe that the field is nearly constant in Euclidean time until it suddenly endures a rapid transition from its false vacuum to the escape point beyond the barrier. This change occurs around the moment $\tilde{\tau}<0$ corresponding to the center of the kink:
 \begin{equation}
 \label{equ:center_solution_toy_homogeneous}
 \sigma(\tilde{\tau})=0 \Rightarrow \tilde{\tau}= -\frac{2C}{v \sqrt{ \lambda }}.
 \end{equation}
 \end{enumerate}

Now that we have explicitly computed the semiclassical exponents in both cases, we rewrite them in terms of dimensionless parameters. First we introduce $\alpha=\frac{\lambda v^4}{ \epsilon}$ such that the approximation (\ref{equ:condition_small_epsilon}) of small $\epsilon$ reduces to $\alpha \gg 8$. Another useful quantity is the ratio $\beta = \frac{R_H}{R^{\star}}$ between the size of the homogeneous bounce and the radius of the thin-wall bubble. It is straightforward to see that the semiclassical decay exponents reduce to 
\begin{equation}
\label{equ:decay_exponent_dimensionless}
S_{B,TW}= \frac{8\pi^2}{3}\frac{\alpha^3}{\lambda} \quad \quad S_{B,hom}=\frac{128 \pi}{9}\frac{\alpha^3 \beta^3}{\lambda}.
\end{equation}
From this parametrization, we can give a few remarks. First, the thin-wall exponent is generally large and leads to a highly suppressed decay probability. Indeed, for the minimal acceptable value $\alpha \sim 10$ and for $\lambda$ of order one, $S_{B,TW}$ is already of order four. On the other hand, there is more freedom for the homogenous factor $S_{B,hom}$, because of the parameter $\beta$ entering in it. When $\beta=1$, namely when the bounces are of the same size, we observe that the two types of decay have a similar probability to occur. Since the radius $R^{\star}$ of the thin-wall bubble is very large, it is reasonable to also consider a homogeneous tunneling taking place in a smaller volume. In that case, the decay probability would be more significant. For instance, when $\alpha=10$, $\beta=0.1$ and $\lambda=1$, we find $S_{B,hom} \approx 45$ and $e^{-S_{B,hom}}\approx 10^{-20}$. We shall make use of this discussion to investigate how the backreaction of particle production modifies these results.

\subsection{\label{sub:homogeneous_decay}Backreaction during the homogeneous bounce}
We consider the production of $\phi$ particles during the homogeneous tunneling process. As described in Sec. \ref{sub:particle_production}, the key parameter to introduce is the coupling $m^2(\sigma)$ between the two fields $\sigma$ and $\phi$. Since the homogeneous bounce (\ref{equ:solution_field_equation_toy_homogeneous}) is almost a step function, we approximate this coupling in the following way:
\begin{equation}
 \label{equ:step_mass_coupling_homogeneous}
 m^2(\sigma)=m^2(\tau)= \left\{ \begin{array}{lll}
 m^2_{-} & \text{ for } & \tau < \tilde{\tau}\\
 m^2_{+} &\text{ for } & \tau > \tilde{\tau}
 \end{array}\right.,
 \end{equation}
where $m_{\pm}$ are the masses of the $\phi$ particles in the true and false vacua, respectively. We remind that the transition occurs at the instant $\tilde{\tau}$ given by Eq. (\ref{equ:center_solution_toy_homogeneous}) and that $\tilde{\tau}<0$.

In this model, Rubakov's prescription to compute the matrix $D$ can be performed easily. According to the spherical background geometry, the indices of this matrix are given by the discrete radial momentum $k=\frac{n}{R_H}$ ($n=0,1,2,\ldots$) and the usual angular momentum $(l,m)$ ($0\leq l \leq n$, $ -l \leq m \leq l$). After a computation detailed in Appendix \ref{app:homogeneous}, similar to the one performed by Rubakov (\cite{Rubakov:1984}, Sec. 4), we note that $D$ is diagonal and does not depend on $(l,m)$. Omitting these indices, the diagonal part of this matrix reads
 \begin{equation}
 \label{equ:D_matrix_solution_homogeneous}
 D_{nn}(\tau=0)=\frac{\omega_+-\omega_-}{\omega_+ + \omega_-}e^{2\omega_+ \tilde{\tau}},
 \end{equation}
where 
 \begin{equation}
 \label{equ:omega_energy_homogeneous}
 \omega_{\pm}= \sqrt{k^2 + m^2_{\pm}}=\sqrt{\frac{n^2}{R_H^2} + m^2_{\pm}}.
 \end{equation}
It is worth directly rewriting this expression in terms of dimensionless quantities. In addition to the two parameters $\alpha$, $\beta$ already introduced above, we define $\delta_{\pm}= R_H m_{\pm}$. From the previous equation for $D_{nn}$ and expression (\ref{equ:center_solution_toy_homogeneous}) for $\tilde{\tau}$, it is straightforward to show that 
\begin{equation}
\label{equ:D_matrix_homogeneous_dimensionless}
D_{nn}=\frac{\sqrt{n^2+\delta_+^2}-\sqrt{n^2+\delta_-^2}}{\sqrt{n^2+\delta_+^2}+\sqrt{n^2+\delta_-^2}}e^{-2 \frac{\sqrt{n^2+\delta_+^2}}{\tilde{\alpha}\beta} },
\end{equation}
where $\tilde{\alpha}:=\frac{\alpha}{\text{arctanh}\left(1-\sqrt{\frac2{\alpha}}\right)}$.

In this way, the backreaction is entirely determined by the four parameters ($\alpha$, $\beta$, $\delta_{\pm}$). We can now compute the correction factor entering in Eq. (\ref{equ:decay_probability_backreaction_final}). Since $D$ is diagonal, the logarithm of $\mathbbm{1}-D^2$ is just the logarithm of each diagonal element and we get
\begin{equation}
\label{equ:backreaction_factor_homogeneous}
\frac12 \Tr \ln\left(\mathbbm{1}-D^2\right)=\frac12 \sum_{n=0}^{+\infty} \sum_{l=0}^{n} \sum_{m=-l}^{l} \ln \left( 1- D_{nn}^2 \right) =\frac12 \sum_{n=0}^{+\infty} (n+1)^2 \ln \left( 1- D_{nn}^2 \right).
 \end{equation}
For consistency with our discussion in Sec. \ref{sub:interpretation}, we explicitly check that this series is convergent. Indeed, its asymptotic expansion is given by
 \begin{equation}
 \label{equ:asymptotic_expansion_backreaction}
 \frac12 (n+1)^2 \ln \left( 1- D_{nn}^2 \right)
  \xrightarrow[n \to \infty]{}
  \left[ - \frac{1}{32 n^2} \left( \delta_+^2-\delta_-^2\right)^2 + O\left( \frac{1}{n^3}\right) \right] e^{-4\frac{n}{\tilde{\alpha} \beta}},
 \end{equation}
meaning that the backreaction is exponentially suppressed at high momentum. It is also clear that the dominant part in Eq. (\ref{equ:backreaction_factor_homogeneous}) comes from the $n=0$ term and so the computation of this leading contribution will already tell us if the backreaction is negligible or significant compared to $S_{B,hom}$. For simplicity, we restrict this analysis to two limiting cases, namely a small and a large mass difference.

 \begin{itemize}
 \item Small mass difference: $ \quad \Delta \delta^2 := \delta_+^2-\delta_-^2 \ll \delta_+^2$.\\
 In this limit, the $n=0$ term in the series (\ref{equ:backreaction_factor_homogeneous}) becomes
 \begin{equation}
 \label{equ:small_mass_expansion}
\frac12 \ln \left( 1- D_{00}^2 \right) = - \frac1{32} \left(\frac{\Delta \delta^2}{\delta_+^2} \right)^2 e^{-4 \frac{\delta_+}{\tilde{\alpha}\beta} } + O\left(\left(\frac{\Delta \delta^2}{\delta^2_+}\right)^3\right).
 \end{equation}
and we directly conclude that this contribution is small. Indeed, the exponential is bounded by $1$ and $\frac{\Delta \delta^2}{\delta_+^2} \ll 1$ because of the small mass difference assumption. Thus we conclude that the backreaction is negligible in front of values of $S_{B,hom}$ which are typically bigger than order one.
 \item Large mass difference: $\delta_- \ll \delta_+$. \\
 Under this assumption, we obtain
\begin{equation}
\label{equ:large_mass_expansion}
\frac12 \ln \left( 1- D_{00}^2 \right) = \frac12 \ln \left(1- e^{-4 \frac{\delta_+}{\tilde{\alpha}\beta} } \right) + O\left( \frac{\delta_{-}}{\delta_+}\right),
 \end{equation} 
where the leading term corresponds to directly take $\delta_-=0$. In contrast with the previous case, this contribution can be arbitrarily large when $\frac{\delta_+}{\tilde{\alpha}\beta}$ becomes small. Thus we should investigate if the backreaction could become of the order of  $S_{B,hom}$ for some reasonable values of these parameters. It turns out that the expression (\ref{equ:large_mass_expansion}) grows very slowly because of the logarithm. Hence the backreaction would unlikely be significant in front of large values of $S_{B,hom}$. For instance for $(\alpha, \beta)=(10,1)$, we saw that $S_{B,hom}=O(10^4)$ and so $\delta_+$ should be as small as many thousand orders of magnitude for the backreaction to be non-negligible.\\
However the situation is more interesting for smaller decay exponents. When $(\alpha, \beta)=(10,0.1)$ we found $S_{B,hom} \approx 45$ and in this case the correction (\ref{equ:large_mass_expansion}) becomes significant for acceptable values of $\delta_+$. Explicitly, we have
\begin{equation}
\label{equ:evaluation_correction_homogeneous}
\left.-\frac12 \ln \left( 1- D_{00}^2 \right)\right|_{(\alpha, \beta, \delta_+)=\left(10,0.1,\left\{10^{-1}, 10^{-4}, 10^{-10}\right\}\right)} \approx  \left\{0.8,\ 4.1,\ 11.2\right\}.
 \end{equation}
In terms of the decay probability, this corresponds to corrections respectively of order $e^{0.8}=O(1)$, $e^{4.1}=O(10^2)$ and $e^{11.2}=O(10^5)$, in front of $e^{-S_B}\approx 10^{-20}$. It confirms that in this case the backreaction enhances the decay rate by some orders of magnitude.
 \end{itemize}

Although we have not investigated the entire range of parameters of this toy model, the two above cases already give us useful information. The first remark is that large semiclassical decay exponents would generally be insensitive to the production of particles. However, we were also able to exhibit a choice of parameters leading to weaker values of $S_{B,hom}$ which are significantly modified by the particle backreaction.

\subsection{\label{sub:tw_backreaction}Backreaction during the thin-wall bounce}
We want to perform the same kind of analysis as above but during the nucleation process of the thin-wall bubble. Since the bounce solution (\ref{equ:solution_field_equation_thin_wall}) is again almost a step function, the system is also well described by the two quantities $m_+$, $m_-$ corresponding to the masses of the $\phi$ particles in the background of the true and false vacua, respectively.

In contrast to the homogeneous case, the matrix $D$ is not diagonal. A tractable way to compute it is the use of the weak particle production limit described in Sec. \ref{sub:weak_particle}. Under this approximation, we saw in Sec. \ref{sub:interpretation} that the backreaction correction is given by one half of the total number of created particles as $\frac12 \Tr D^2$. Fortunately, Rubakov (\cite{Rubakov:1984}, Sec. 5) already computed this quantity since he was interested in the number of created particles in this model. We present some details of his calculation in Appendix \ref{app:D_matrix_thin_wall} for consistency and directly give his results here. He proved that the main contribution to the number of produced particles comes from the case $m_- \ll (R^{\star})^{-1} \ll m_+$ and from particles with radial momentum $m_- \ll p \ll m_+$ and angular momentum $l=m=0$. In these conditions, he found that $N=\Tr D^2=O(10^{-2})$. It means that the number of particle is weak (at the best roughly one particle produced per $100$ bubbles) whatever the values of the parameters.

Remembering that $S_{B,TW}$ in Eq. (\ref{equ:decay_exponent_dimensionless}) is at least of order $O(10^4)$, it is clear that the backreaction factor $\frac12 \Tr D^2$ is completely negligible.

\section{Conclusion and outlook}
This paper offers a new approach to compute the backreaction of particle production on the decay rate of a false vacuum. We explicitly derived a formula which corrects the usual semiclassical decay probability, in the case of a tunneling field coupled to a scalar environment in flat spacetime. Starting from Rubakov's formalism describing the spectrum of created particles, the main idea of this work was to integrate out this external bath of particles using the reduced density matrix prescription. 

As a first consequence, we found that the correction factor is ultraviolet finite. Hence its computation does not require any renormalization techniques in contrast with the calculation of one-loop quantum corrections. We also showed that scalar particle production enhances the decay probability in the context of this formalism. It may be interpreted as the sign that the dissipation of the tunneling field into the environment is compensated by the external fluctuations. Another important observation is the fact that the backreaction is given by one half of the total number of created particles, in the approximation that they are weakly produced. In other words, Rubakov's prescription gives directly both the spectrum of particles and their backreaction in this limit.

These general remarks would not be of particular importance if the backreaction were always negligible compared to the semiclassical decay rate. That is why we explicitly computed this effect for a toy model potential. We found a negligible impact in the case of the thin-wall bubble nucleation. However, when the field decays homogeneously in a finite volume, we computed a significant correction for a reasonable choice of parameters. Therefore, it would be interesting to apply this formalism to more interesting systems, especially to the Higgs potential of the Standard Model. Of course, a complete analysis of the SM requires extending this formalism to subjects that we should investigate in a subsequent work, including spinor fields, vector fields and eventually gravitational effects.

\begin{acknowledgments}
I would like to thank the Sydney University Particle Physics Group for its hospitality during my master project and the elaboration of this paper. I am especially grateful to Archil Kobakhidze for his supervision and for useful discussion. I also thank Mikhail Shaposhnikov for his final revision of my master thesis. This work was partially supported by the Australian Research Council.
\end{acknowledgments}

\appendix

\section{\label{app:combinatorial}Combinatorial argument}
\begin{figure}
\centering
\includegraphics[scale=0.8]{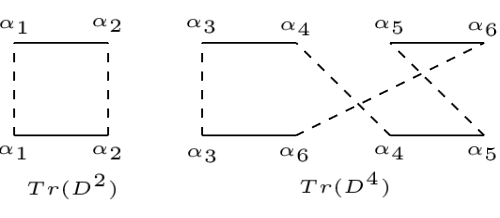}
\caption{Diagram corresponding to the case $k=3$ and the permutation $(1)(2)(3)(465)$. In terms of partition, it corresponds to $j_1=1, j_2=1, j_3=0$. The sum over $\alpha_1,\ldots \alpha_6$ gives the result $\Tr\left(D^2\right)^{j_1} \Tr \left(D^4\right)^{j_2} \Tr \left(D^6\right)^{j_3}$.}
\label{fig:diagram_k_3}
\end{figure}

We explicitly derive Eq. (\ref{equ:ch4_coefficient_C_k_combinatoric}) for the coefficients $C_{\left\{ j_i\right\}_k}$ entering into the derivation of the backreaction factor. We remind that they are defined in order to satisfy the following equation for a fixed integer $k$:
\begin{equation}
\label{equ:equality_sum_over_permutation_app}
\sum_{\left\{ \alpha \right\}_{2k}} \sum_{\lambda \in S_{2k}} \prod_{i=1}^k D_{\alpha_{2i-1}\alpha_{2i}} D_{\alpha_{\lambda(2i-1)}\alpha_{\lambda(2i)}} = \sum_{\left\{ j_i\right\}_k} C_{\left\{ j_i\right\}_k} \prod_{i=1}^{k} \Tr\left(D^{2i}\right)^{j_i}.
\end{equation}
The left-hand side of this expression becomes clearer if we introduce a diagrammatic representation. For each $D_{\alpha_{2i-1} \alpha_{2i}}$, we draw a horizontal solid segment with the two end points $\alpha_{2i-1}$ and $\alpha_{2i}$. Then we horizontally align all these factors to picture the product over $i$. We do exactly the same thing for the $D_{\alpha_{\lambda(2i-1)}\alpha_{\lambda(2i)}}$ but we align them below the first product. The last step is to connect with a dashed line each end point $\alpha_{i}$ with the corresponding end point $\alpha_{\lambda(j)}$ where $\lambda(j)=i$. There is such a diagram for each permutation $\lambda \in S_{2k}$ and we give one example in Fig. \ref{fig:diagram_k_3}. 

When we take the sum over $\left\{ \alpha \right\}_{2k}$, each connected block of $2i$ solid segments gives a factor of the form $\Tr \left( D^{2i} \right)$. It explains the form of the right-hand size of Eq. (\ref{equ:equality_sum_over_permutation_app}). Indeed, each diagram can be decomposed in $j_1$ blocks of two solid lines, $j_2$ blocks of four solid lines and so on where the $j_i$ form a partition of $k$ as explained in the main text. Of course, it is clear that many permutations give the same partition and this is actually what we want to compute. Let us consider a given partition $\left\{ j_i\right\}_k$. To compute $C_{\left\{ j_i\right\}_k}$, we proceed in two steps. First, we forget the internal structure of each block in the diagrams described above and we compute the number of ways, say $A_{\left\{ j_i\right\}_k}$, in which we can arrange these blocks under the constraint given by the partition. In the second step, we compute the number of internal structures, say $B_{\left\{ j_i\right\}_k}$, corresponding to each arrangement. 

\begin{itemize}
\item Expression for $A_{\left\{ j_i\right\}_k}$

Since there are $k$ segments on each row of each diagram, we can arrange them in $(k!)^2$ different ways. We still have to divide by the correct overcounting factor. First, a permutation between the blocks of same size does not change the structure, so we can divide by $\prod_{i=1}^k j_i!$. Second, consider a block of size $(2i)$. Since we do not care about its internal structure, we can permute the segments of the upper row and the segments of the lower row independently. It gives a factor $(i!)^2$ for each block and so the final expression is  

\begin{equation}
\label{equ:expression_for_A_app}
A_{\left\{ j_i\right\}_k} = \frac{(k!)^2}{\prod_{i=1}^k j_i! (i!)^{2j_i}}.
\end{equation}

\item Expression for $B_{\left\{ j_i\right\}_k}$

Consider again a block of size $(2i)$. We want to find the number of ways to connect each upper end point to a lower end point in order to keep the block in a single connected piece. We proceed as follows. The first upper point can be linked to one of the $2i$ lower points. Then the lower point lying on the same segment can be connected to only $2i-2$ upper point in order not to close the cycle. We again have $2i-2$ possibilities for the next upper points and so on and so forth. So the total contribution of the block is $(2i)(2i-2)(2i-2)(2i-4)(2i-4)\ldots 1 = (2i)!!(2i-2)!!$. The double factorial of an even integer can be simplified as $(2i)!!=2^i i!$. So the total contribution from all the blocks becomes
\begin{equation}
\label{equ:expression_for_B_app}
B_{\left\{ j_i\right\}_k}=\prod_{i=1}^{k} \left( (2i)!!(2i-2)!!\right)^{j_i}= \prod_{i=1}^{k} \left( \frac{2^{2i} (i!)^2}{2i}\right)^{j_i}= 2^{2k} \prod_{i=1}^{k} \frac{(i!)^{2j_i}}{(2i)^{j_i}},
\end{equation}
where we have used $\prod_{i=1}^{k} 2^{2i j_i}=2^{2k}$ since the $j_i$ are a partition of $k$.
\end{itemize}

It is clear that the second expression is the same for each one of the $A_{\left\{ j_i\right\}_k}$ arrangements of the blocks. It means that the final coefficient is just the product of Eq. (\ref{equ:expression_for_A_app}) and Eq. (\ref{equ:expression_for_B_app}):
\begin{equation}
\label{equ:app_coefficient_C_k_combinatoric}
C_{\left\{ j_i\right\}_k}= \frac{(k!)^2 2^{2k}}{\prod_{i=1}^{k} j_i! (2i)^{j_i}}.
\end{equation}


\section{\label{app:homogeneous} \texorpdfstring{$D$}{D} matrix for the homogeneous bounce}
We explicitly derive Eq. (\ref{equ:D_matrix_solution_homogeneous}) for the matrix $D$ during the homogeneous decay, according to Rubakov's prescription. In that case, Eq. (\ref{equ:xi_differential_equations}) for the wave functions becomes
 \begin{equation}
 \label{equ:xi_differential_equation_homogeneous_app}
 \left[\partial^2_i  +\left(\omega_{\pm}^{\alpha}\right)^2 - m^2_{\pm}  \right]\xi_{\pm}^{\alpha}=0,
 \end{equation}
 where $\xi_{-}^{\alpha}$ (resp. $\xi_{+}^{\alpha}$) corresponds to $\tau < \tilde{\tau}$ (resp. $\tau > \tilde{\tau}$). Taking into account the spherical background geometry of the process, the solutions of this equation are given by
 \begin{equation}
 \label{equ:solution_xi_homogeneous_decay_app}
 \xi_{\pm}^{nlm}= \frac{1}{\sqrt{2 \omega_{\pm} R_H^3}}Y_{nlm}\left(\frac{\mathbf{x}}{R_H}\right) \quad n=0,1,\ldots  \quad 0\leq l \leq n \quad -l \leq m \leq l,
 \end{equation}
 where $Y_{nlm}$ are the real harmonics on the unit $3$-sphere and the prefactor ensures the normalization condition (\ref{equ:xi_normalization}). The momentum of the particle is discrete and given by $k=\frac{n}{R_H}$ which implies
 \begin{equation}
 \label{equ:omega_energy_homogeneous_app}
 \omega_{\pm}= \sqrt{k^2 + m^2_{\pm}}=\sqrt{\frac{n^2}{R_H^2} + m^2_{\pm}}.
 \end{equation}

Thanks to the homogeneity of the bounce, the functions $h_{\alpha}(\tau,x)$ can be written as $h_{nlm}(\tau, \mathbf{x})=h_n(\tau) Y_{nlm}\left(\frac{\mathbf{x}}{R_H}\right)$ where $h_n(\tau)$ satisfies
 \begin{equation}
 \label{equ:euclidean_equation_h_homogeneous_app}
  \left[- \frac{\partial^2 }{\partial \tau^2} + m^2(\tau) + \frac{n^2}{R_H^2} \right]h_{n}(\tau)=0.
 \end{equation}
As $m^2(\tau)$ is constant in the two regions $\tau < \tilde{\tau}$ and $\tau> \tilde{\tau}$, a solution exponentially decreasing at $\tau \rightarrow - \infty$ is simply given by
 \begin{equation}
 \label{equ:solution_h_tau_homogeneous_app}
 h_n(\tau)= \left\{ \begin{array}{lll}
 e^{\omega_{-} \tau} & \text{ for } & \tau < \tilde{\tau}\\
 A_n e^{\omega_{+} \tau} + B_n e^{- \omega_{+} \tau} &\text{ for } & \tau > \tilde{\tau}
 \end{array}\right.,
 \end{equation}
 where the two coefficients $A_n$, $B_n$ are determined from the continuity conditions at $\tilde{\tau}$:
 \begin{equation}
 \label{equ:A_B_coefficient_homogeneous_app}
 A_n = \frac{\omega_{+} +\omega_{-}}{2 \omega_{+}} e^{-(\omega_{+}-\omega_{-}) \tilde{\tau}} \quad B_n = \frac{\omega_{+} -\omega_{-}}{2 \omega_{+}} e^{(\omega_{+}+\omega_{-}) \tilde{\tau}}.
 \end{equation}
 
 The two matrices $V$ and $Z$ defined by (\ref{equ:matrices_V_Z}) can now easily be computed at $\tau> \tilde{\tau}$:
 \begin{equation}
 \label{equ:V_Z_matrice_homogeneous_app}
 \begin{array}{l}
 V^{+}_{nn',ll',mm'}(\tau)=\delta_{nn',ll',mm'}\left(g_n(\tau)\omega_{+}-\partial_{\tau}g_n(\tau) \right)= 2 B_n e^{-\omega_+ \tau} \ \delta_{nn',ll',mm'}, \\
 Z^{+}_{nn',ll',mm'}(\tau)=\delta_{nn',ll',mm'}\left(g_n(\tau)\omega_{+}+\partial_{\tau}g_n(\tau) \right)= 2 A_n e^{\omega_+ \tau} \ \delta_{nn',ll',mm'}.
 \end{array}
 \end{equation}
 As expected, they are diagonal and do not depend on the angular momentum $(l,m)$. Omitting these indices, the diagonal elements of $D$ at $\tau> \tilde{\tau}$ are
 \begin{equation}
 \label{equ:D_matrix_solution_homogeneous_appendix}
 D_{nn}(\tau)=\frac{B_n}{A_n}e^{-2\omega_+ \tau} \quad \Rightarrow \quad  D_{nn}(0)=\frac{B_n}{A_n}=\frac{\omega_+-\omega_-}{\omega_+ + \omega_-}e^{2\omega_+ \tilde{\tau}}.
 \end{equation}


\section{\label{app:D_matrix_thin_wall}\texorpdfstring{$D$}{D} matrix for the thin-wall bounce}
We briefly show how the matrix $D$ and the number of created particles are computed in the case of the thin-wall bounce. We use the weak particle production limit described in Sec. \ref{sub:weak_particle} and follow the derivation performed by Rubakov (\cite{Rubakov:1984}, Sec. 5). As we consider the nucleation of a symmetric bubble, we can work in spherical coordinates and parametrized the wave functions $\xi^{\alpha}_{\tau}$ as
\begin{equation}
\label{equ:wave_function_xi_radial_app}
\xi_{\tau}^{plm}(\mathbf{x}) = \frac{1}{r \sqrt{\omega_{\tau}^{plm}}}\eta_{\tau}^{plm}(r) Y_{lm}\left(\frac{\mathbf{x}}{r}\right),
\end{equation}
where $Y_{lm}$ are the real spherical harmonics. The label $\alpha$ corresponds now to the radial momentum $p$ of the particle and the angular momentum $(l,m)$.  Eq. (\ref{equ:xi_differential_equations}) and normalization (\ref{equ:xi_normalization}) for $\xi^{\alpha}_{\tau}$ become for $\eta_{\tau}^{plm}$:
\begin{equation}
\label{equ:wave_equation_eta_radial_app}
\left[ - \frac{d^2}{dr^2}+ \frac{l (l+1)}{r^2} + m^2(r,\tau) \right]\eta_{\tau}^{plm}= \left(\omega_{\tau}^{plm}\right)^2 \eta_{\tau}^{plm} ,
\end{equation}
\begin{equation}
\label{equ:normalization_eta_radia_app}
\int_{0}^{+\infty} dr \, \eta_{\tau}^{plm} \eta_{\tau}^{p'l'm'} = \delta(p-p')\delta_{ll'}\delta_{mm'}.
\end{equation}
With this parametrization and omitting the indices $l,m$, Eq. (\ref{equ:matrix_D_weak_limit}) for $D$ becomes
\begin{equation}
\label{equ:matrix_D_nucleation_lowest_order_eta_app}
D_{pp'}(\tau=0) = \int_{-\infty}^0 d\tau e^{W_p(\tau) +W_{p'}(\tau)} \frac{\mathcal{D}(p,p',\tau)}{\omega_{\tau}^p +\omega_{\tau}^{p'}}
\end{equation}
where
\begin{equation}
\label{equ:matrix_d_curl_lowest_order_app}
\mathcal{D}(p,p',\tau)= \int_{0}^{+\infty} dr \frac{\eta_{\tau}^p}{\sqrt{2 \omega_{\tau}^{p}}} \frac{\eta_{\tau}^{p'}}{ \sqrt{2\omega_{\tau}^{p'}}} \frac{\partial m^2}{\partial \tau} \quad \text{and} \quad  W_p(\tau)=\int_{0}^{\tau} d\tau' \omega_{\tau'}^p.
\end{equation}

We now specify the form of the parameter $m^2$ for the thin-wall bounce configuration. As already explained, this parameter reduces to a step function and we can parametrize it as
 \begin{equation}
 \label{equ:ch4_step_thin_wall_bounce_x_tau_app}
 m^2(r, \tau) = \left\{ \begin{array}{lll}
 m_{-}^2  & \text{ if } & \tau < -R^{\star} \\
 m_{+}^2 + (m_-^2-m_+^2) \Theta(r-r_{\star}(\tau)) & \text{ if }  &  -R^{\star} \leq \tau \leq 0\\
 \end{array}\right.,
 \end{equation}
 where $ r_{\star}(\tau)=\sqrt{R^{\star 2}-\tau^2}$ is the size of the Euclidean growing bubble and $\Theta$ is the Heaviside step function. It directly gives us the time derivative of the mass coupling entering in Eq. (\ref{equ:matrix_d_curl_lowest_order_app}):
 \begin{equation}
 \label{equ:ch4_derivative_mass_coupling_app}
 \frac{\partial m^2}{\partial \tau} = - \frac{\tau}{r_{\star}} (m_+^2-m_-^2) \delta (r-r_{\star}),
 \end{equation}
meaning that the particles are produced at the wall of the bubble. According to the form of $m^2$, the solutions $\eta_{\tau}^p$ of Eq.(\ref{equ:wave_equation_eta_radial_app}) at each instant $\tau$ admit an analytical expression in terms of Bessel functions. Their integration in Eq. (\ref{equ:matrix_d_curl_lowest_order_app}) can only be performed analytically under some simplifications. For the case which provides the dominant contribution to the number of particles, namely when $m_- \ll (R^{\star})^{-1} \ll m_+$, $m_- \ll p \ll m_+$ and $l=0$, it is possible to show that
\begin{equation}
\label{equ:D_curl_approx_tw_app}
\mathcal{D}_{l=0}(p,p',\tau)= -\frac{\tau}{\pi r_{\star}}(pp')^{1/2},
\end{equation}
and thus
\begin{equation}
\label{equ:number_particle_approx_tw_app}
\begin{split}
N_{l=0} &=\frac{1}{\pi^2} \int_{0}^{+\infty} dp dp' \int_{-R^{\star}}^0d\tau d\tau' e^{(p+p')(\tau+\tau')} \frac{pp'}{(p+p')^2}\frac{\tau \tau'}{r_{\star}(\tau) r_{\star}(\tau')} \\ &=\frac{1}{12 \pi^2}=O(10^{-2})
\end{split}
\end{equation}
which is the result quoted in Sec. \ref{sub:tw_backreaction}. For more details, like the the treatment of the other cases  or the explicit form of the functions $\eta_{\tau}^p$, we directly refer to \cite{Rubakov:1984}.

\bibliography{backreaction_biblio}

\end{document}